\documentclass[11pt]{article}
\usepackage[latin9]{inputenc}
\usepackage{geometry}
\usepackage{amsmath}
\usepackage{amssymb}

\makeatletter
\usepackage{graphicx}
\usepackage{amsthm}

\makeatother

\begin{document}

\title{\textbf{A Comparative Study of magneto-thermo-elastic wave propagation
in a finitely conducting medium under thermoelasticity of type I,
II, III}}

\date{{*}Dr.Rakhi Tiwari, Prof. J. C. Mishra}

\maketitle
\begin{center}
\emph{{*}Department of Mathematics, Babasaheb Bhimrao Ambedakar Bihar
University}\\
\emph{ Department of Mathematics, Indian Institute of Engineering,
Science and Technology}\\
 \emph{{*}E-mail: rakhibhu2117@gmail.com}
\par\end{center}

\noindent \textbf{Abstract:}\ The present work is concerned with
the propagation of electro-magneto-thermoelastic plane waves of assigned
frequency in a homogeneous isotropic and finitely conducting elastic
medium permeated by a primary uniform external magnetic field. We
formulate our problem under the theory of Green and Naghdi of type-III
(GN-III) to account for the interactions between the elastic, thermal
as well as magnetic fields. A general dispersion relation for coupled
waves is deduced to ascertain the nature of waves propagating through
the medium. Perturbation technique has been employed to obtain the
solution of dispersion relation for small thermo-elastic coupling
parameter and identify three different types of waves. We specially
analyze the nature of important wave components like, phase velocity,
specific loss and penetration depth of all three modes of waves. We
attempt to compute these wave components numerically to observe their
variations with frequency. The effect of presence of magnetic field
is analyzed. Comparative results under theories of type GN-I, II and
III have been presented numerically in which we have found that the
coupled thermoelastic waves are un-attenuated and non-dispersive in
case of Green-Naghdi-II model which is completely in contrast with
the theories of type-I and type-III. Furthermore, the thermal mode
wave is observed to propagate with finite phase velocity in case of
GN-II model, whereas the phase velocity of thermal mode wave is found
to be an increasing function of frequency in other two cases. We achieve
significant variations among the results predicted by all three theories.

\noindent \vspace{5bp}


\vspace{0.02\columnwidth}

\noindent \textbf{1.\ Introduction\vspace{0.01\columnwidth}
}

\noindent \ \ The development of active materials have drawn the
serious attention of researchers towards the interactions between
magnetic, thermal and mechanical fields in a thermo-elastic solid
in presence of magnetic field. The magnetic field originating inside
nuclear reactors and the extremely high temperature affect their design
and operations {[}1{]}. This terminology is acknowledged as the theory
of magneto-thermo-elasticity. This topic has several applications
in various fields like, geophysics, optics, acoustics, plasma physics,
damping of acoustic wave in magnetic fields and other related topics
involving sensing and actuation. The materials with electric-elastic
coupling are used as ultrasonic transducers and micro-actuators, while
materials with thermal and electric coupling have been used for thermal
imaging devices. The concept of elasto-magnetic coupling effects in
materials have applications in health monitoring of civil structures.
Basically, the coupled theory of electro-magneto-thermoelasticity
is the combination of two different disciplines namely theory of electro-magnetism
and theory of thermoelasticity. A systematic presentation of governing
equations along with the uniqueness and reciprocity theorem for linear
thermo-electro-magneto-elasticity can be found in the article by Li
{[}2{]}.

\noindent \ \ The study of plane wave propagation in thermo-elastic
material in presence of external magnetic field has been the topic
of high interest during last few decades. Paria{[}3{]}, Wilson {[}4{]}
and Purushotama{[}5{]} have used the classical theory of thermoelasticity
(Biot{[}6{]}) along with the electromagnetic theory to characterize
harmonically time dependent plane waves of assigned frequency in a
homogeneous, isotropic and unbounded solid. But unfortunately, the
classical theory of thermoelasticity by Biot {[}6{]} is unrealistic
from a physical point of view, particularly in problems like those
concerned with sudden heat inputs, because it exhibits infinite speed
of propagation of thermal waves. Hence, to account for demerits of
this theory, some theories have been developed as extended theory
of classical thermoelasticity. The theories given by Lord and Shulman
{[}7{]} and Green and Lindsay {[}8{]} are two theories that have been
extensively employed by researchers as generalized theories of thermoelasticity.
Subsequently, Green and Naghdi {[}9,10,11{]} developed three theories
of thermoelasticity which we refer to in the present work as Green
Naghdi -I, II and III (GN-I, GN-II and GN-III) theory of thermoelasticty.
The linearized version of GN-I theory is identical to the classical
theory of thermoelasticity. GN-II theory accounts for finite speed
of thermal wave, but with no energy dissipation. This implies that
undamped waves of second sound effect is identified. However GN-III
theory does not follow GN-I and GN-II but shows GN-I \& GN-II theories
as its special cases.

\noindent \ \ Nayfeh and Nemat-Nesser {[}12{]} and later on, Agarwal
{[}13{]} reported a detailed study on electro-magneto-thermoelastic
plane waves in solids in the context of generalized thermoelasticity
theory with the effects of thermal relaxation parameters. Roychoudhuri
{[}14,15{]} and Roychoudhuri and Debnath {[}16,17{]} studied propagation
of magneto-thermoelastic plane waves in rotating thermoelastic media
permeated by a primary uniform magnetic field using generalized heat
conduction equation of Lord and Shulman {[}7{]}. Chandrasekharaiah
{[}18{]} investigated the harmonic plane wave propagation in an unbounded
medium by employing GN-II theory of thermoelasticity. Puri and Jordan
{[}19{]} and later on, Kothari and Mukhopadhyay {[}20{]} used GN-III
theory to explain the propagation of thermoelastic plane waves. Roychoudhuri
and Banerjee (Chattopadhyay) {[}21{]} discussed magnetoelastic plane
waves in rotating media under thermoelasticity of type-II model. Das
and Kanoria {[}22{]} also worked on magneto-thermo-elastic waves in
a medium with infinite conductivity in the context of GN-III theory.
Abbas and Alzahrani {[}23{]} studied Green-Naghdi Model in a 2D problem
of a mode I crack in an Isotropic Thermoelastic Plate. Moreover Ezzat,
El-Karamany and El-Bary {[}24{]} presented Two-temperature theory
in Green\textendash{}Naghdi thermoelasticity with fractional phase-lag
heat transfer. Tiwari and Mukhopadhyay {[}25{]} studied electromagnetic
wave propagation in GN-II model of thermoelasticity.

\noindent \ \ At present, our motive of this work is to investigate
the propagation of magneto-thermo-elastic plane waves of assigned
frequency in an infinite, homogeneous, isotropic, thermally and electrically
conducting solid in the context of GN-I, GN-II and GN-III theory of
thermoelasticity. We have taken the media having finite conductivity
permeated by a primary uniform external magnetic field. The basic
governing equations are derived and a more general dispersion relation
is obtained in GN-III theory of thermoelasticity. 'Perturbation technique'
has been enforced to solve the problem analytically. Three various
types of plane waves are identified by solving the general dispersion
relation analytically and various physical characterizations of plane
waves are deduced. The problem is illustrated with numerical results
of various wave characterizations and the nature of waves of different
modes are critically analyzed. Since GN-I and GN-II an be obtained
as a special cases of GN-III model of thermoelasticity. Therefore,
the nature of electro-magnetic harmonic plane waves have been characterized
under the theories of Green and Naghdi of types I, II and III (GN-I,
GN-II and GN-III) using numerical results and conclusions are explained
to highlight the specific features of the present investigation.

\vspace{0.02\columnwidth}

\noindent \textbf{2.\ Problem formulation and basic governing equations\vspace{0.01\columnwidth}
}

\noindent \ \ For our present study, an infinite, homogeneous, isotropic,
thermally and electrically conducting solid permeated by a primary
magnetic field $\vec{B_{0}}=(B_{1},B_{2},B_{3})$ is considered. The
media is characterized by the density $\rho$ and Lame's elastic constants
$\lambda$ and $\mu$.

\noindent \ \ Using a fixed rectangular cartesian coordinate system
\emph{(x,y,z}) and employing the thermoelasticity theory of Green
and Naghdi {[}{[}9{]}-{[}11{]}{]}, the equations of motion and the
equation of heat conduction in the presence of magnetic field in the
absence of external body force (mechanical) and heat sources can be
represented in the following manner:

\noindent Equation of motion:

\noindent 
\begin{equation}
\mu\nabla^{2}\vec{u}+(\lambda+\mu)\vec{\nabla}(\vec{\nabla}.\vec{u})+\vec{J}\times\vec{B}-\nu\vec{\nabla}\theta=\rho\ddot{\vec{u}}
\end{equation}
Equation of heat conduction {[}9{]}:

\noindent 
\begin{equation}
K^{*}\nabla^{2}\theta+K\nabla^{2}\dot{\theta}=\rho C_{v}\ddot{\theta}+\nu T_{0}\ddot{u_{i,i}}
\end{equation}
where $\vec{J}$ is the current density vector and $\vec{J}\times\vec{B}$
is the electromagnetic body force (Lorentz force).

\noindent \ \ Here, $\vec{u}$ is the displacement field, $\theta$
is the temperature above reference temperature $T_{0}$. Total magnetic
field $\vec{B}=\vec{B_{0}}+\vec{b}$ is assumed to be small so that
the products with $\vec{b}$ and $\dot{\vec{u}}$ and their derivatives
can be neglected for the linearization of the field equations. $\vec{b}=(b_{x},b_{y},b_{z})$
is the perturbed magnetic field. $\lambda,\mu$ are Lame's constants
and $\nu=(3\lambda+2\mu)\alpha_{t}$, where $\alpha_{t}$ is the coefficient
of thermal expansion of the solid. Dots denote the derivatives with
respect to the time \emph{t}. $C_{v}$ is the specific heat of the
solid at constant volume, $K^{*}$ is the thermal conductivity rate
and $K$ is the thermal conductivity of the medium.

\noindent \ \ \textbf{\emph{Notice that if we put $K*=0$ in equation
(2) i.e. the thermal conductivity rate is absent, then the equation
is acknowledged by the heat conduction equation for GN-I theory of
thermoelasticity and if we substitute $K=0$ in equation (2) i.e.
the thermal conductivity rate is absent, we obtain the heat conduction
equation of GN-II theory of thermoelasticity.}}

\noindent \ \ Due to the presence of magnetic field inside the medium,
equation of motion needs to be supplemented by generalized Ohm's law
in a continuous medium with Max well\textquoteright{}s electromagnetic
field equations.\\
 Max well\textquoteright{}s equations (where the displacement current
and charge density are neglected) are given by\\
\begin{equation}
\vec{\nabla}\times\vec{H}=\vec{J}
\end{equation}
\begin{equation}
\vec{\nabla}\times\vec{E}=-{\frac{\partial\overrightarrow{B}}{\partial t}}
\end{equation}
where $\vec{B}=\mu_{e}\vec{H}$ and $\mu_{e}$ is the magnetic permeability.

\begin{equation}
\vec{\nabla.}\vec{B}=0
\end{equation}
Generalized Ohm's law is given by 
\begin{equation}
\vec{J}=\sigma[\vec{E}+{\frac{\partial u}{\partial t}}\times\vec{B}]
\end{equation}
where $\sigma$ is the electrical conductivity, ${\frac{\partial u}{\partial t}}$
is the particle velocity of the medium. Here the small effect of temperature
gradient on $\vec{J}$ is neglected.

\vspace{0.02\columnwidth}
\textbf{3.\ Dispersion relation and its analytical solution\vspace{0.01\columnwidth}
}

We are assuming that plane waves are propagating towards \emph{x}-direction.
Due to this all the field quantities are proportional to $e^{i(kx-\omega t)}$,
where \emph{k} is the wave number and $\omega$ is the angular frequency
of plane waves. Here, we have assumed that $\omega$ is real and \emph{k}
may be complex quantity such that $Im(k)\leq0$ for waves to be physically
realistic. In the context of above consideration, we can write all
the quantities in the following manner:
\begin{description}
\item [{$\vec{u}=(u,v,w)=(u_{0},v_{0},w_{0})e^{i(kx-\omega t)}$,$\ \theta=\theta_{0}e^{i(kx-\omega t)}$;}]~
\item [{$\vec{E}=(E_{x},E_{y},E_{z}),\vec{\ J}=(j_{1},j_{2},j_{3})e^{i(kx-\omega t)}$;}]~
\item [{$\vec{b}=(b_{x},b_{y},b_{z})=(b_{1},b_{2},b_{3})e^{i(kx-\omega t)}$}]~
\end{description}
\noindent where $u_{0}$, $v_{0}$, $w_{0}$; $j_{1}$, $j_{2}$,
$j_{3}$, $b_{1}$, $b_{2}$, $b_{3}$, $\theta_{0}$ are constants.\\
\\
\ \ With the help of Max well\textquoteright{}s relations, we achieve
the following relation:
\begin{equation}
div\overrightarrow{b}=0
\end{equation}
\ \ Above relation implies that $b_{x}=0.$ Using equation (3) we
obtain 
\begin{equation}
\mu_{e}\overrightarrow{J}=\vec{\nabla}\times\overrightarrow{b}.
\end{equation}

\noindent which is in agreement with the following value of $\vec{J}$

\begin{equation}
\vec{J}=[0,-{\frac{ikb_{z}}{\mu_{e}}},{\frac{ikb_{y}}{\mu_{e}}}]
\end{equation}

\noindent and
\begin{equation}
\vec{J}\times\vec{B_{0}}=[-{\frac{ik(b_{z}B_{3}+b_{y}B_{2})}{\mu_{e}}},{\frac{ikb_{y}B_{1}}{\mu_{e}}},{\frac{ikb_{z}B_{1}}{\mu_{e}}}]
\end{equation}
Thus the term $\overrightarrow{J}\times\overrightarrow{B}$ can be
replaced by the term $\overrightarrow{J}\times\overrightarrow{B_{0}}$.\\
\ \ Substituting values of the quantities $\vec{u}$ and $\theta$
in the equation of heat conduction , we achieve the following relation:

\begin{equation}
\theta=\alpha u_{0}
\end{equation}

\noindent where 
\begin{equation}
\alpha=\frac{i\nu\theta_{0}k\omega^{2}}{K^{*}k^{2}-iKk^{2}\omega-\rho C_{v}\omega^{2}}
\end{equation}
We obtain $\vec{E}=(E_{x},E_{y},E_{z})=(E_{x},{\frac{\omega b_{z}}{k}},-{\frac{\omega b_{y}}{k}})$\\
 \ \ Now, making use of the field quantities $\vec{u}$, $\vec{J}$
and $\vec{E}$ and comparing both sides of the modified Ohm's law
(in which $\vec{B}$ is replaced by $\vec{B_{0}}$), we achieve the
following three relations:

\noindent 
\begin{equation}
\sigma[E_{x}-i\omega(qB_{3}-rB_{2})]=0
\end{equation}
\begin{equation}
\sigma[\frac{\omega b_{z}}{k}-i\omega(rB_{1}-pB_{3})]=-\frac{ikb_{z}}{\mu_{e}}
\end{equation}
\begin{equation}
\sigma[-\frac{\omega b_{y}}{k}-i\omega(pB_{2}-qB_{1})]=\frac{ikb_{y}}{\mu_{e}}
\end{equation}
\ \ Further, substituting the values of field quantities $\overrightarrow{u}$,$\overrightarrow{J}$,$\vec{B_{0}}$,$\vec{\theta}$
in the equation of motion, we achieve the following relations:

\begin{equation}
u_{0}(-\rho\omega^{2}+(\lambda+2\mu)k^{2}+i\nu\alpha k)+\frac{ik}{\mu_{e}}(b_{3}B_{3}+b_{2}B_{2})=0
\end{equation}

\begin{equation}
v_{0}(-\rho\omega^{2}+\mu k^{2})-\frac{ik}{\mu_{e}}(b_{2}B_{1})=0
\end{equation}

\begin{equation}
w_{0}(-\rho\omega^{2}+\mu k^{2})-\frac{ik}{\mu_{e}}(b_{3}B_{1})=0
\end{equation}
 \ \ Equations (14) and (15) can be rewritten as:

\noindent 
\begin{equation}
u_{0}\sigma(i\omega B_{3})+w_{0}(-\sigma i\omega B_{1})+b_{3}[\frac{ik}{\mu_{e}}-\frac{\sigma\omega}{k}]=0
\end{equation}
 
\begin{equation}
u_{0}\sigma(-i\omega B_{2})+v_{0}(\sigma i\omega B_{1})-b_{2}[\frac{ik}{\mu_{e}}+\frac{\sigma\omega}{k}]=0
\end{equation}
\ \ Equations {[}(16)-(20){]} constitute a system of five equations
with five unknowns namely $u_{0}$, $v_{0}$, $w_{0}$, $b_{2}$,
$b_{3}$.

\noindent \ \ We can make further assumptions: we have taken that
$\vec{b}$ is directed along y-axis and we consider $w_{0}=0$ provided
that $(\mu k^{2}-\rho\omega^{2}\neq0)$ so that $b_{3}=0$ (equation
(18)). Hence, applied and perturbed magnetic field are taken as $(\overrightarrow{B_{1}},\overrightarrow{B_{2}},0)$
and $(0,\overrightarrow{b_{2}},0)$, respectively. On applying the
above assumptions in the equations {[}(16)-(20){]}, all equations
reduce into three following homogeneous equations having three unknowns
$u_{0}$, $v_{0}$ and $b_{2}$ as\\
\begin{equation}
u_{0}[-\rho\omega^{2}+(\lambda+2\mu)k^{2}+i\nu\alpha k]+\frac{ikB_{2}b_{2}}{\mu_{e}}=0
\end{equation}
\begin{equation}
v_{0}(-\rho\omega^{2}+\mu k^{2})-\frac{ikB_{1}b_{2}}{\mu_{e}}=0
\end{equation}
\begin{equation}
u_{0}(\sigma i\omega B_{2})-v_{0}\sigma(i\omega B_{1})+b_{2}(\frac{ik}{\mu_{e}}+\frac{\sigma\omega}{k})=0
\end{equation}
\ \ In order to find the solution for the system of equations (21),
(22) and (23) for $u_{0},$$v_{0}$and $b_{2}$ equations, we can
write

\noindent 
\[
\left|\begin{array}{ccc}
-\rho\omega^{2}+(\lambda+2\mu)k^{2}+i\nu\alpha k & 0 & \frac{ikB_{2}}{\mu_{e}}\\
\\
0 & \mu k^{2}-\rho\omega^{2} & \frac{-ikB_{1}}{\mu_{e}}\\
\\
i\sigma\omega B_{2} & 0 & (\frac{ik}{\mu_{e}}+\frac{\sigma\omega}{k})
\end{array}\right|=0
\]
\ \ Further, we assume that the initial magnetic field is directed
towards y axis i.e. $\vec{B_{0}}=(0,B_{2},0)$. Therefore $B_{1}=0$
in the determinent. For simplifying the system for achieving the solution,
we introduce the following non- dimensional quantities in above: \\
$\chi=\frac{\omega}{\omega^{*}}$, $\xi=\frac{kc_{1}}{\omega^{*}},\ $$\epsilon_{H}=\frac{\omega^{*}\nu_{H}}{{c_{1}^{2}}}$\\
$\epsilon_{\theta}=\frac{T_{0}\nu^{2}}{\rho^{2}C_{v}{c_{1}^{2}}},\ \nu_{H}=\frac{1}{\mu_{e}\sigma},$\\
$k_{1}=\frac{K^{*}}{\rho C_{v}c_{1}^{2}},\ k_{2}.\frac{K\omega^{*}}{\rho C_{v}c_{1}^{2}}$\\
 $c_{1}=\sqrt{\frac{(\lambda+2\mu)}{\rho}}$, $c_{2}=\sqrt{\frac{\mu}{\rho}}$\\
where $c_{1}$ is the longitudinal elastic wave velocity and $c_{2}$
is the transverse elastic wave velocity. We further assume that\\
 $s=\frac{c_{2}}{c_{1}}$, $R_{H}=\frac{B_{2}^{2}}{\rho c_{1}^{2}\mu_{e}}$.
\\
where $R_{H}$ is the magnetic pressure number. Substituting the value
of $\alpha$ from eqn. (12) and employing all dimensionless quantities
in the expansion of the above determinant, we find the following dispersion
relation:
\begin{align}
(s^{2}\xi^{2}-\chi^{2})[(\xi^{2}-\chi^{2})(\xi^{2}k_{1}-\chi^{2})-ik_{2}\xi^{2}\chi(\xi^{2}-\chi^{2})-\epsilon_{\theta}\xi^{2}\chi^{2}](\chi+i\xi^{2}\epsilon_{H})\nonumber \\
+R_{H}\chi\xi^{2}[(\xi^{2}k_{1}-\chi^{2})-ik_{2}\xi^{2}\chi]=0\,\,\,
\end{align}
\ \ The first part $(s^{2}\xi^{2}-\chi^{2})=0$ in (24) corresponds
to a transverse elastic wave which is clearly found to be uncoupled
by thermal and magnetic field. Hence, we take the second part of (24)
as\\
\begin{equation}
[(\xi^{2}-\chi^{2})(\xi^{2}k_{1}-\chi^{2})-ik_{2}\xi^{2}\chi(\xi^{2}-\chi^{2})-\epsilon_{\theta}\xi^{2}\chi^{2}](\chi+i\xi^{2}\epsilon_{H})+R_{H}\chi\xi^{2}[(\xi^{2}k_{1}-\chi^{2})-ik_{2}\xi^{2}\chi]=0
\end{equation}
\ \ Equation (25) is clearly identified as the dispersion relation
for coupled thermal-dilatational-electrical waves propagating in the
medium in the present context. With the help of this relation, we
will characterize the behaviour of plane waves propagating in the
present context. We will specially concentrate on the analysis of
the important wave characterizations like, phase velocity, specific
loss and penetration depth. For solving equation (25), we attempt
to find the perturbation solution of the dispersion equation for small
values of thermoelastic coupling constant$\epsilon_{\theta}$. Therefore,
substituting $\epsilon_{\theta}$= 0 in the dispersion relation, we
achieve the following solutions:

\begin{equation}
\xi^{2}=A\chi^{2}
\end{equation}

where $A=\frac{k_{1}+ik_{2}\chi}{k_{1}^{2}+k_{2}^{2}\chi^{2}}=a_{1}+ib_{1}$

\begin{equation}
[(\xi^{2}-\chi^{2})(\chi+i\xi^{2}\epsilon_{H})+R_{H}\chi\xi^{2}]=0
\end{equation}
\vspace{0.01\columnwidth}
\ \ Equation (27) being an equation of degree 4, we consider the
roots of above equation (27) as $\pm\alpha_{1}$and $\pm\alpha_{2}$. 

where 
\begin{equation}
\alpha_{1,2}^{2}=\frac{i\chi^{2}\epsilon_{H}-\chi(1+R_{H})\pm\left\{ \chi^{2}(1+R_{H})^{2}-\chi^{4}\epsilon_{H}^{2}-2\chi^{3}(1+R_{H})\epsilon_{H}i+4i\epsilon_{H}\chi^{3}\right\} ^{\frac{1}{2}}}{2i\epsilon_{H}}
\end{equation}
\ \ Which implies:

\begin{equation}
\alpha_{1,2}^{2}=\frac{i\chi^{2}\epsilon_{H}-\chi(1+R_{H})\pm(a+ib)}{2i\epsilon_{H}}
\end{equation}

Therefore

\begin{equation}
\alpha_{1}^{2}=\frac{i\chi^{2}\epsilon_{H}-\chi(1+R_{H})+(a+ib)}{2i\epsilon_{H}}
\end{equation}

\begin{equation}
\alpha_{2}^{2}=\frac{i\chi^{2}\epsilon_{H}-\chi(1+R_{H})-(a+ib)}{2i\epsilon_{H}}
\end{equation}

where

$a=\sqrt{\frac{\chi^{2}(1+R_{H})^{2}-\chi^{4}\epsilon_{H}^{2}+\chi^{2}[\{(1+R_{H})^{2}+\chi^{2}\epsilon_{H}^{2}-4\chi\epsilon_{H}\sqrt{R_{H}}\}\{(1+R_{H})^{2}+\chi^{2}\epsilon_{H}^{2}+4\chi\epsilon_{H}\sqrt{R_{H}}\}]^{\frac{1}{2}}}{2}}$

$b=\sqrt{\frac{-\chi^{2}(1+R_{H})^{2}+\chi^{4}\epsilon_{H}^{2}+\chi^{2}[\{(1+R_{H})^{2}+\chi^{2}\epsilon_{H}^{2}-4\chi\epsilon_{H}\sqrt{R_{H}}\}\{(1+R_{H})^{2}+\chi^{2}\epsilon_{H}^{2}+4\chi\epsilon_{H}\sqrt{R_{H}}\}]^{\frac{1}{2}}}{2}}$

\vspace{0.01\columnwidth}
\ \ As we are attempting to find the perturbation solution for small
values of thermoelastic coupling constant $\epsilon_{\theta}.$ Therefore
$\xi^{2}$ can be written in the following form-

\begin{equation}
\xi_{1}^{2}=\alpha_{1}^{2}+n_{1}\epsilon_{\theta}+O(\epsilon_{\theta}^{2})
\end{equation}

\begin{equation}
\xi_{2}^{2}=\alpha_{2}^{2}+n_{2}\epsilon_{\theta}+O(\epsilon_{\theta}^{2})
\end{equation}

\begin{equation}
\xi_{3}^{2}=A\chi^{2}+n_{3}\epsilon_{\theta}+O(\epsilon_{\theta}^{2})
\end{equation}
\ \ Substituting equations (32), (33) and (34) in equation (25)
and comparing the lowest power of $\epsilon_{\theta}$ on both sides
of equation and neglecting the terms of $O(\epsilon_{\theta}^{2})$,
we obtain the following solution:

\begin{equation}
\xi_{1}^{2}=\alpha_{1}^{2}[1+A\epsilon_{\theta}\chi^{2}\frac{(\chi+i\alpha_{1}^{2}\epsilon_{H})}{G_{1}}]
\end{equation}

\begin{equation}
\xi_{2}^{2}=\alpha_{2}^{2}[1+A\epsilon_{\theta}\chi^{2}\frac{(\chi+i\alpha_{2}^{2}\epsilon_{H})}{G_{2}}]
\end{equation}

\begin{equation}
\xi_{3}^{2}=A\chi^{2}[1+\frac{A\epsilon_{\theta}\chi^{3}(1+iA\chi\epsilon_{H})}{(A-1)(\chi^{3}+iA\epsilon_{H}\chi^{4})+R_{H}A\chi^{3}}]
\end{equation}

where 
\begin{equation}
G_{1,2}=((\chi+i\alpha_{1,2}^{2}\epsilon_{H})(\alpha_{1,2}^{2}-\chi^{2})+\chi R_{H}\alpha_{1,2}^{2})+(\alpha_{1,2}^{2}-\chi^{2})((1+R_{H})\chi+2i\alpha_{1,2}^{2}\epsilon_{H}-i\epsilon_{H}\chi^{2})
\end{equation}
Clearly, the above expressions for $\xi_{i},i=1,2,3$ (with $Im(\xi_{i}\leq0)$
correspond to three modes of plane waves propagating inside the medium. 

\textbf{4.\ Analytical expressions of various components of magneto-thermo-elastic
plane wave\vspace{0.01\columnwidth}
}\ \ In order to calculate several components of waves like, phase
velocity, specific loss and penetration depth we use the following
formulae:\vspace{0.01\columnwidth}

\textbf{Phase velocity:} 
\begin{equation}
V_{1,2,3}=\frac{\chi}{Re[\xi_{1,2,3}]}
\end{equation}

\textbf{Specific loss:}
\begin{equation}
S_{1,2,3}=4\pi|\frac{Im[\xi_{1,2,3}]}{Re[\xi_{1,2,3}]}|
\end{equation}

\textbf{Penetration depth:}
\begin{equation}
D_{1,2,3}=\frac{1}{|Im[\xi_{1,2,3}]|}
\end{equation}
From the solutions obtained as above, we can clearly observe that
there are three modes of waves which are dissimilar to each other.
We denote the first wave as modified (quasi-magneto) elastic (dilatational)
wave, the second one as modified (quasi-magneto) thermal wave and
the third one as (quasi-magneto) electrical wave. It is observed that
both the elastic and thermal mode wave are influenced by the thermoelastic
coupling constants $\epsilon_{\theta}$, magneto-elastic coupling
constant $\epsilon_{H}$, as well as magnetic pressure number $R_{H}$.
We will specially concentrate on these two modes. Since it is very
difficult to characterize the nature of waves from analytical solutions
obtained above. Therefore, the various components of magneto-thermoelastic
plane wave like, phase velocity, specific loss and penetration depth
for all three kinds of waves regarding GN-I, GN-II and GN-III theory
of thermoelasticity by numerical results. 

\vspace{0.02\columnwidth}

\textbf{5.\ Numerical results}: \textbf{Analysis of various features
of plane wave\vspace{0.01\columnwidth}
}

From the analytical results obtained above, we conclude that three
various modes of waves have been extracted from the coupled dispersion
relation which are named as modified (quasi-magneto) elastic (dilatational)
wave, modified (quasi-magneto) thermal wave and (quasi-magneto) electrical
wave. In order to illustrate the analytical solution and to have a
critical analysis of the nature of waves with the variation of frequency,
we will make an attempt to show the variations of various wave components
of the identified waves in three different models of thermolelasticity
with the help of numerical results. We will also assess the limiting
behaviour of wave components.

We make an attempt to represent the plane wave characterizations numerically
with the help of computational work by using programming on Mathematica.
Copper material has been chosen for the purpose of numerical evaluations.
The physical data for our problem are taken as\\
 $\epsilon_{T}=0.0168$\\
 $\lambda=7.76\times10^{10}Nm^{-2}$\\
 $\mu=3.86\times10^{10}Nm^{-2}$\\
 $\mu_{e}=4*10^{-7}NA^{-2}$\\
 $\sigma=5.7*10^{7}Sm^{-1}$\\
$\omega^{*}=1.72*10^{11}sec^{-1}$\\
$\rho=8954Kgm^{-3}$\\
\ \ We assume: $k_{1}=1$and $k_{2}=1$.

We will specially analyze the physical parameters of waves like, phase
velocity, specific loss and penetration depth. Using the formulae
given by equations (39)-(41), we compute these components of waves
of different modes and display our results in various Figures. Our
analytical work is devoted to GN-III theory of magneto thermoelasticity
for a finitely conducting medium. Furthermore, in order to make a
comparison between GN-I, GN-II \& GN-III models, we carry out computational
work for these two special cases too (GN-I and GN-II). Due to this
reason, we have considered the variations in the propagation of waves
for GN-I and GN-III model of thermoelasticity in each figure. Since
we are obtaining a very different nature of waves in GN-II theory
of thermoelasticity as compare to GN-I and GN-II theories of thermoelasticity;
therefore, nature of waves in GN-II model of thermoelasticity has
been represented in seperate plots. We have taken the values of the
non dimensional thermal conductivity rate i.e. $k_{1}=1,$ $k_{2}=1$
in GN-III model. $k_{1}=0$ represents the case of GN-I model. $k_{2}=0$
represents the case of GN-II model. In each Figure, the thin solid
line is used for $k_{1}=0$(GN-I), thick dashed line is for $k_{1}=1$
and $k_{2}=1$ (GN-III), We find that the trend of all the wave components
of the third mode wave, i.e., electric mode wave are almost similar
in the contexts of GN-I, GN-II and GN-III model. However, a prominent
difference in the results under GN-I model, GN-II model and GN-III
model is indicated for the quasi-magneto elastic wave and quasi-magneto
thermal mode wave. We highlight several specific features arisen out
of the numerical results in the following sections:

\vspace{0.02\columnwidth}

\textbf{5.1.\ Analysis of phase velocity}:\vspace{0.01\columnwidth}

\ \ Using the formula given by (39), we compute the phase velocity
of all three modes of waves. Figs.1(\emph{a}) and 1(\emph{b}) display
the variation of phase velocity of modified quasi-magneto dilatational
wave and phase velocity of modified quasi-magneto thermal wave, respectively
in GN-I and GN-III model of thermoelasticity. Fig. 1(c) represents
the variation of phase velocities of quasi-magneto dilatational wave
and quasi-magneto thermal wavein GN-II model of thermoelasticity.
We observe from Fig.1(\emph{a}) that the starting from a intial value,
phase velocity of quasi-magneto dilatational wave propagates in increasing
mode and after giving a extreme value it becomes constant very soon.
In GN-I model, after achieving a local maximum value, wave goes down
and becomes constant. But in GN-III model, wave is achieving the local
maximum value and at local maximum value it becomes constant. We obtain
a slight difference between the extreme values as well as constant
values in phase velocity in both models. Surprisingly in all cases
each figure goes towards a constant limiting value which is nearer
to 1. 

Fig. 1(b) describe the variations of phase velocity of quasi-magneto
thermal wave in GN-I and GN-III models of thermoelasticity. In the
present context, waves are propagating from a constant limiting value
greater than 0 when frequency tends to zero value and then starts
increasing with respect to frequency and goes towards infinity. Here,
we are obtaining the similar nature of wave under GN-I and GN-III
models of thermoelasticity. 

Here, we are achieving a significant result that a prominent difference
in the phase velocity of quasi-magneto dilatational wave in GN-I and
GN-III models; however mode of propagation for quasi-magneto dilatational
wave is similar in nature ion GN-I and GN-III models of thermoelasticity
i.e. quasi-magneto thermal mode wave exhibits a remarkable difference
between two models (GN-I and GN-III) as compare to quasi-magneto dilatational
wave.

As we have already stated above that when $k_{2}=0$, we obtain the
equations for Green and Naghdi-II (GN-II) model of magneto-thermoelasticity.
Fig. 1(c) exhibits the behaviour of phase velocities of quasi-magneto
dilatational wave and quasi-magneto thermal wave in GN-II model of
thermoelasticity. Here we obtain that phase velocity of quasi-magneto
dilatational wave as well as of quasi-magneto thermal mode wave show
almost similar trend of variation and have constant limiting value.
Phase-velocity of thermal wave is greater that the phase velocity
of elastic mode wave and both of them reaches to constant value as
we increase the frequency. In this case, we achieve completely different
nature of quasi-magneto dilational wave and quasi-magneto thermal
wave in comparison to the results predicted by two models GN-I and
GN-III. We have already identified that the speed of modified (quasi
magneto) thermal wave increases with the increases of frequency in
cases of GN-I and GN-III models and goes towards infinity, but it
is constant in GN-II model and this is the special feature of GN-II
model of magneto-thermoelasticity that the speed of magneto-thermal
wave is always finite in this case. 

In the present case, wave is unaffected with the values of $k_{1}$and
$k_{2}$ and the predictions of GN-I, GN-II and GN-III model is almost
similar in nature. Therefore, Fig. 4(\emph{a}) represents the behaviour
of phase velocity of quasi-magneto electrical wave in all models .
Here, phase velocity of waves increases rapidly with respect to frequency. 

\vspace{0.02\columnwidth}
\textbf{ 5.2.\ Analysis of specific loss}:\vspace{0.01\columnwidth}

Specific loss is defined as the loss of energy per stress cycle as
defined by formula (40). Figures 2(\emph{a}) and 2(\emph{b}) exhibit
the variations of specific loss of quasi-magneto dilatational mode
wave and quasi-magneto thermal wave, respectively. In quasi-magneto
dilatational wave, initially specific loss is 0 but it starts increasing
with respect to frequency and giving a maximum value of specific loss
it starts decreasing and goes towards 0 value. This maximum value
is different for GN-I and GN-III however the difference is not prominent. 

Nature of quasi-magneto thermal wave is completely different from
quasi-magneto dilatational wave. It starts from a initial value and
after achieving a maximum value which is nearer to 1; it becomes constant.
Here we are observing that energy loss is greater in quasi-magneto
dilatational wave as compare to quasi-magneto dilatational wave. There
is one significant point is that in quasi-magneto dilatational wave,
maximum value of specific loss is greater in GN-I as compare to GN-III
which implies that GN-III model exhibits better results as compare
to GN-I model.

Fig.2(c) represents the combined study of specific loss for Quasi-magneto-dilatational
wave and Quasi-Magneto-thermal wave in GN-II model of thermoelasticity.
Here the picture is complely changed as compare to GN-I and GN-III
model of thermoelasticity. Here, the value of loss is very less nearer
to 0. As we have studied that there is no specific loss in GN-II model
of thermoelasticity without magnetis field But here we see that magnetic
field is effective which means that in presence of magnetic field,
the ideal nature of GN-II variates and we are finding wave-energy-loss.
Nature of loss for both type of waves are almost similar.

Fig.4(b) expresses the behaviour of specific loss for quasi-magneto-electrical
wave. Here we are achieving a constant value of loss and the value
is appromately 1. Furthermore, we obtain the result that there is
no variation in the behaviour of quasi-magneto electrical wave for
GN-I, GN-II and GN-III model of thermoelasticity i.e. we obtain the
same result in all three models.

\vspace{0.02\columnwidth}
\textbf{5.3.\ Analysis of penetration depth}:\vspace{0.01\columnwidth}

The behaviour of penetration depth of quasi-magneto dilatational wave
and quasi-magneto thermal wave can be seen from Figs. 3(\emph{a,b}),
respectively which show that the penetration depth of quasi-magneto
dilatational wave decreases from infinite value as the frequency increases
and reaches to constant limiting value nearer to119. But we observe
significant differences among the plots of this field for different
GN-I and GN-III. We see that in GN-III model, penetration depth of
quasi-magneto dilatational wave decreases from infinite value and
after giving a local minimum it starts increases and become constant
at the limiting value nearer to 119 However in GN-I model, penetration
depth decreases from infinite and becomes constant that in GN-III,
is more prominent for lower frequency values, although in all cases,
it finally reaches constant value nearer to 119. Here, there is no
significicant difference between the constant limiting values under
two models (GN-I and GN-III).

For quasi-magneto thermal wave, penetration depth decreases ferom
infinite and goes down to a very low constant value nearer to 0. THere
is no significant difference between two models GN-I and GN-III.

Fig.3(c) displays the penetration depth of Quasi-magneto-dilatational
wave and Quasi-Magneto-thermal wave in GN-II model. Here we are getting
completely different results as compare to GN-I and GN-III. In the
present context, penetration depth decreases from infinite and becoms
constant at a very high value nearer to in power of $10^{8}$which
is very large as compare to the constant limiting value in case of
GN-I and GN-III. Here we can interpretate that the wave is nearer
to the ideal behaviour because as we have stated that there is no
enery loss in GN-II without magnetic field but in the presence of
magnetic field, the nature of waves variates slightly in GN-II and
shows that the nature of waves variates from ideality and giving a
little loss in wave-energy. Moreover there is no the prominent difference
in the nature of variation in depth for both qausi-magneto dilational
wave and quasi-magneto thermal wave.

Penetration depth of quasi-magneto electrical wave exhibits rapidly
decreasing behaviour with respect to frequency and becomes constant
with the finite value in both the GN-I and GN-III models of thermoelasticity
(see Fig. 4(c)) and there is no significant differences in the results
predicted by GN-I and GN-III models.

\vspace{0.02\columnwidth}
\textbf{7.\ Summary and observations:\vspace{0.01\columnwidth}
}

In the present work, dispersion relation solutions for the plane wave
propagating in a magneto-thermoelastic media with finite electrical
conductivity have been determined by employing Green and Naghdi theory
of thermoelasticity of type-III. We have made a comparative study
of GN-I, GN-II and GN-III theory of thermoelasticity in presence of
an external magnetic field. From the derived dispersion relation solution,
transverse and longitudinal plane waves are investigated. We find
that transverse mode elastic wave is uncoupled from the thermal and
magnetic field; Further a general dispersion relation associated to
the coupled dilatational-thermal and electrical wave is identified
and we make attempt to extract three different modes of waves from
this coupled dispersion relation. These waves are identified as quasi-magneto
dilatational wave, quasi-magneto thermal wave and quasi-magneto electrical
wave. The quasi-magneto-electrical wave is found to have similar variation
under GN-I, GN-II, GN-III theory. However, significant differences
are obtained in other two modes, namely quasi-magneto dilatational
and quasi-magneto thermal mode wave predicted by three different models.
Hence, we pay attention to these three modes and analyze various wave
components like, phase velocity, specific loss and penetration depth.
The behaviour of the wave components in limiting cases of frequency
values have been investigated with the help of graphical plots. Various
features are highlighted. It is believed that this study would be
useful due to its various applications in different areas of physics,
geophysics etc. The most highlighted features of the present investigation
can be summarized as follows:
\begin{enumerate}
\item Significant resemblance and non- resemblance among the results under
GN-I, GN-II and GN-III theory of thermoelasticity have been identified. 
\item The phase velocity of thermal mode wave is found to be an increasing
function of frequency under GN-I and GN-III models.
\item Quasi- magneto dilatational and thermal mode waves propagate faster
in the theory of type GN- I in comparison to GN-III theory of thermoelasticity.
However, phase velocity of quasi magneto-electric wave is unaffected
whether we employ GN-I, GN-II or GN-III theory. Quasi-magneto dilatational
and thermal wave is found to be nearer to non-dispersive in GN-II
model, however due to presence of magnetic field, there is very less
specific loss in waves and that is why we are getting a constant limiting
value of penetration depth however the constant value is very high
but not equal to infinite in the context of GN-II theory of thermoelasiticity.
\end{enumerate}
Penetration depth has a less finite value in the case of GN-I and
GN-III theory of thermoelasticity. However in GN-II, we see that penetration
depth for both waves namely, quasi-magneto dilatational wave and quasi-magneto
thermal wave is very high since waves are propagating with constant
speed and there is a very less specific energy loss of waves. This
is a very distinct feature of GN-II model.
\begin{enumerate}
\item In view of above points, we can conclude that for coupled magneto-thermoalstic
problem, GN-II model exhibits realistic behaviour in comparison to
GN-I and GN-III models w.r.t. phase velocity of thermal wave, but
when we analyze the behaviour of penetration depth, we find that predictions
of GN-I and GN-III theory is more realistic as compared to GN-II model
as we obtain in this case a very high nearer to infinite penetration
depth which is also physically unrealistic prediction by GN-II model.\end{enumerate}

\end{document}